\documentclass[12pt]{article}
\begin{document}
\title{THE ORIGIN OF MASS, SPIN AND INTERACTION}
\author{B.G. Sidharth\\
International Institute for Applicable Mathematics \& Information Sciences\\
Hyderabad (India) \& Udine (Italy)\\
B.M. Birla Science Centre, Adarsh Nagar, Hyderabad - 500 063 (India)}
\date{}
\maketitle
\begin{abstract}
We argue that a non commutative geometry at the Compton scale is at the root of mass, Quantum Mechanical spin and QCD and electromagnetic interactions. It also leads to a reconciliation of linearized General Relativity and Quantum Theory.
\end{abstract}
\vspace{5 mm}
\begin{flushleft}
\end{flushleft}
\section{Introduction}
Modern Fuzzy Spacetime and Quantum Gravity approaches deal with a non differentiable spacetime manifold. In the latter approach there is a minimum spacetime cut off, which, as shown nearly sixty years ago by Snyder leads to what is nowadays called a non commutative geometry, a feature shared by the Fuzzy Spacetime also \cite{snyder,amati,uof,kempf,madore,mota}. The new geometry is given by
\begin{equation}
[dx^\mu , dx^\nu ] \approx \beta^{\mu \nu} l^2 \ne 0\label{e14}
\end{equation}
While equation (\ref{e14}) is true for any minimum cut off $l$, it is most interesting and leads to physically meaningful relations including a rationale for the Dirac equation and the underlying Clifford algebra, when $l$ is at the Compton scale (Cf.ref.\cite{uof}). In any case given (\ref{e14}), the usual invariant line element, 
\begin{equation}
ds^2 = g_{\mu \nu} dx^\mu dx^\nu\label{e15}
\end{equation}
has to be written in terms of the symmetric and nonsymmetric combinations for the product of the coordinate differentials. That is the right side of Equation (\ref{e15}) would become
$$\frac{1}{2} g_{\mu \nu} \left[\left(dx^\mu dx^\nu + dx^\nu dx^\mu\right) + \left(dx^\mu dx^\nu - dx^\nu dx^\mu\right)\right],$$
In effect we would have
\begin{equation}
g_{\mu \nu} = \eta_{\mu \nu} + kh_{\mu \nu}\label{e16}
\end{equation}
So the noncommutative geometry introduces an extra term, that is the second term on the right side of (\ref{e16}). It has been shown in detail by the author that (\ref{e14}) or (\ref{e15}) lead to a reconciliation of electromagnetism and gravitation and lead to what may be called an extended gauge formulation \cite{nc1,nc2,ijmpe,afdb}.\\
The extra term in (\ref{e16}) leads to an energy momentum like tensor but it must be stressed that its origin is in the non commutative geometry (\ref{e14}). All this of course is being considered at the Compton scale of an elementary particle.
\section{Compton Scale Considerations}
As in the case of General Relativity \cite{ohan,mwt}, but this time remembering that neither the coordinates nor the derivatives commute we have
$$\partial_\lambda \partial^\lambda h^{\mu \nu} - (\partial_\lambda \partial^\nu h^{\mu \lambda} + \partial_\lambda \partial^\mu h^{\nu \lambda})$$
\begin{equation}
-\eta^{\mu \nu} \partial_\lambda \partial^\lambda h + \eta^{\mu \nu}\partial_\lambda \partial_\sigma h^{\lambda \sigma} = - k T^{\mu \nu}\label{e17}
\end{equation}
It must be reiterated that the non commutativity of the space coordinates has thrown up the analogue of the energy momentum tensor of General Relativity, viz., $T^{\mu \nu}$. We identify this with the energy momentum tensor.\\
Remembering that $h_{\mu \nu}$ is a small effect, we can use the methods of linearized General Relativity \cite{ohan,mwt}, to get from (\ref{e17}), 
\begin{equation}
g_{\mu v} = \eta_{\mu v} + h_{\mu v}, h_{\mu v} = \int
\frac{4T_{\mu v}(t-|\vec x - \vec x'|,\vec x')}{|\vec x - \vec x'|}
d^3x'\label{e18}
\end{equation}
It was shown several years ago in the context of linearized General Relativity, that for distances $|\vec{x} - \vec{x'}|$ much greater than the distance $\vec{x'}$, that is well outside the Compton wavelength, we can recover from (\ref{e18}) the electromagnetic potential (Cf.ref.\cite{cu} and references therein). We will briefly return to this point.\\
Let us now see what happens when $|\vec x| \sim |\vec x'|.$ In this case, we have from (\ref{e18}), expanding in a Taylor series about $t$,
\begin{eqnarray}
h_{\mu v} = 4 \int \frac{T_{\mu v}(t,\vec x')}{|\vec x - \vec x'|}d^3 x'+
(\mbox{terms   independent  of}\vec x) + 2 \nonumber \\
\int \frac{d^2}{dt^2} T_{\mu v} (t,\vec x'). |\vec x - \vec x'| d^3 x' +
0(|\vec x - \vec x'|^2)\label{e19}
\end{eqnarray}
The first term gives a Coulombic $\frac{\alpha}{r}$ type interaction except
that the coefficient $\alpha$ is of much greater magnitude as compared to
the gravitational or electromagnetic case, because in an expansion of $(1/|\vec x - \vec x'|),$ all terms are of comparable order. The second term on the right side of (\ref{e19}) is of no dynamical value as it is independent of $\vec{x}$. The third term however is of the form constant $\times r$. That is the potential (\ref{e19}) is exactly of the form of the QCD potential \cite{lee}
\begin{equation}
-\frac{\alpha}{r} + \beta r\label{e20}
\end{equation}
In (\ref{e20}) $\alpha$ is of the order of the mass of the particle as follows from (\ref{e19}) and the fact that $T^{\mu \nu}$ is the energy momentum tensor given by 
\begin{equation}
T^{\mu \nu} = \rho u^\mu u''\label{e21}
\end{equation}
where $u^\mu$ represented the four velocity. Remembering that from (\ref{e14}), we are within a sphere of radius given by the Compton length where the velocities equal that of light, we have equations
\begin{equation}
|\frac{du_v}{dt}| = |u_v|\omega\label{e22}
\end{equation}
\begin{equation}
\omega = \frac{|u_v|}{R} = \frac{2mc^2}{\hbar}\label{e23}
\end{equation}
Alternatively we can get (\ref{e22}) from the theory of the Dirac equation itself \cite{diracpqm}, viz., 
$$\imath \hbar \frac{d}{dt} (u_\imath) = -2mc^2 (u_\imath),$$
Using (\ref{e21}), (\ref{e22}) and (\ref{e23}) we get
\begin{equation}
\frac{d^2}{dt^2} T^{\mu v} = 4 \rho u^\mu u^v \omega^2 = 4 \omega^2 T^{\mu v}\label{e24}
\end{equation}
Equation (\ref{e24}) too is obtained in the Dirac theory (loc.cit). Whence, as can be easily verified, $\alpha$ and $\beta$ in (\ref{e20}) have the correct values required for the QCD potential (Cf. also \cite{cu}). (Alternatively $\beta r$ can be obtained, as in the usual theory by a comparison with the Regge angular momentum mass relation: It is in fact the constant string tension like potential which gives quark confinement and its value is as in the usual theory \cite{phys}).\\
Let us return to the considerations which lead via a non commutative geometry to an energy momentum tensor in (\ref{e17}). We can obtain from here the origin of mass and spin itself, for we have as is well known (Cf.ref.\cite{mwt})
$$m = \int T^{00} d^3 x$$
and via
$$S_k = \int \epsilon_{klm} x^l T^{m0} d^3 x$$
the equation
$$S_k = c < x^l > \int \rho d^3 x \cdot$$
While $m$ above can be immediately and consistently identified with the mass, the last equation gives the Quantum Mechanical spin if we remember that we are working at the Compton scale so that
$$\langle x^l \rangle = \frac{\hbar}{2mc}\cdot$$ 
Returning to the considerations in (\ref{e14}) to (\ref{e17}) it follows that (Cf.ref.\cite{nc1})
\begin{equation}
\frac{\partial}{\partial x^\lambda} \frac{\partial}{\partial
x^\mu} - \frac{\partial}{\partial x^\mu} \frac{\partial}{\partial
x^\lambda} \mbox{goes over to} \frac{\partial}{\partial x^\lambda}
\Gamma^\nu_{\mu \nu} - \frac{\partial}{\partial x^\mu}
\Gamma^\nu_{\lambda \nu}\label{De9c}
\end{equation}
Normally in conventional theory the right side of (\ref{De9c})
would vanish. Let us designate this non vanishing part on the right
by
\begin{equation}
\frac{e}{c\hbar} F^{\mu \lambda}\label{De10c}
\end{equation}
We have shown here that the non commutativity in momentum
components leads to an effect that can be identified with
electromagnetism and in fact from expression (\ref{De10c})
we have
\begin{equation}
A^\mu = \hbar \Gamma^{\mu \nu}_\nu\label{De11c}
\end{equation}
where $A_\mu$ as noted can be identified with the electromagnetic
four potential and the Coulomb law deduced for $|\vec{x}-\vec{x}'|$ in (\ref{e18}) much greater than $|\vec{x}'|$ that is well outside the Compton scale (Cf.ref.\cite{uof} and also ref. \cite{cu}).  To see this in the light
of the usual gauge invariant minimum coupling (Cf.ref.\cite{cu}),
we start with the effect of an infinitesimal parallel displacement
of a vector in this non commutative geometry,
\begin{equation}
\delta a^\sigma = - \Gamma^\sigma_{\mu \nu} a^\mu
dx^\nu\label{De12c}
\end{equation}
As is well known, (\ref{De12c}) represents the effect due to the
curvature and non integrable nature of space - in a flat space,
the right side would vanish. Considering the partial derivatives
with respect to the $\mu^{th}$ coordinate, this would mean that,
due to (\ref{De12c})
\begin{equation}
\frac{\partial a^\sigma}{\partial x^\mu} \to \frac{\partial
a^\sigma}{\partial x^\mu} - \Gamma^\sigma_{\mu \nu}
a^\nu\label{Daa}
\end{equation}
Letting $a^\mu = \partial^\mu \phi,$, we have, from (\ref{Daa})
$$D_{\mu \nu} \equiv \partial_\nu \partial^\mu \to D'_{\mu \nu}\equiv \partial_\nu \partial^\mu - \Gamma^\mu_{\lambda \nu} \partial^\lambda$$
\begin{equation}
= D_{\mu} - \Gamma^\mu_{\lambda \nu} \partial^\lambda\label{Dexx}
\end{equation}
Now we can also write
$$D_{\mu \nu} = (\partial^\mu - \Gamma^\mu_{\lambda \lambda}) (\partial_\nu - \Gamma^\lambda_{\lambda \nu}) + \partial^\mu \Gamma^\lambda_{\lambda \nu} + \Gamma^\mu_{\lambda \lambda} \partial_\nu$$
So we get
$$D_{\mu \nu} - \Gamma^\mu_{\lambda \lambda} \partial_\nu = (p^\mu) (p_\nu)$$
where
$$p^\mu \equiv \partial^\mu - \Gamma^\mu_{\lambda \lambda}$$
Or,
$$D_{\mu \mu} - \Gamma^\mu_{\lambda \lambda} \partial_\mu = (p^\mu) (p_\mu)$$
Further we have
$$D'_{\mu \mu} = D_{\mu \mu} - \Gamma^\mu_{\lambda \mu} \partial^\lambda$$
Thus, (\ref{Dexx}) gives, finally,
$$D'_{\mu \nu} = (p_\mu) (p_\nu)$$
That is we have
$$\frac{\partial}{\partial x^\mu} \to \frac{\partial}{\partial x^\mu} - \Gamma^\nu_{\mu \nu}$$
Comparison with (\ref{De11c}) establishes the required identification.\\
It is quite remarkable that equation (\ref{De11c}) is mathematically identical to Weyl's unified formulation, though this was not originally acceptable because of the adhoc insertion of the electromagnetic potential. Here in our case it is a consequence of the geometry - the noncommutative geometry (Cf.refs.\cite{cu} and \cite{bgsafdb} for a detailed discussion).\\
It was also described in detail how in the usual commutative spacetime the Dirac
spinorial wave functions conceal the noncommutative
character (\ref{e14}) \cite{uof}.\\
Indeed we can verify all these considerations in a simple way as
follows:\\
First let us consider the usual spacetime, in which the
Dirac wave function is given by
$$\psi = \left(\begin{array}{ll}
\chi \\ \Theta
\end{array}\right),$$
where $\chi$ and $\Theta$ are two component spinors. It is well known
that under reflection while the so called positive energy
spinor $\Theta$ behaves normally, on the contrary $\chi \to -\chi , \chi$
being the so called negative energy spinor which comes
into play at the Compton scale \cite{bd}. That is, space is doubly connected. Because of this property as shown in
detail \cite{nc2}, there is now a covariant derivative
given by, in units, $\hbar = c=1$,
\begin{equation}
\frac{\partial \chi}{\partial x^\mu} \to [\frac{\partial}{\partial
x^\mu} - n A^\mu]\chi\label{De12}
\end{equation}
where
\begin{equation}
A^\mu = \Gamma^{\mu \sigma}_{\sigma} = \frac{\partial} {\partial
x^\mu} log (\sqrt{|g|)}\label{De13}
\end{equation}
$\Gamma$ denoting the Christofell symbols.\\
$A^\mu$ in (\ref{De13})is now identified with the
electromagnetic potential, exactly as in Weyl's
theory except that now, $A^\mu$ arises from the bi spinorial
character of the Dirac wave function or the double connectivity of spacetime. In other words, we return to (\ref{De11c}) via an alternative route.\\
What all this means is that the so called ad hoc feature in
Weyl's unification theory is really symptomatic of
the underlying noncommutative spacetime geometry (\ref{e14}).
Given (\ref{e14}) (or (\ref{e16})) we
get both gravitation and electromagnetism in a unified picture, because both are now the consequence of spacetime geometry. We could think that gravitation arises from the symmetric part of the metric tensor (which indeed is the only term if $0(l^2)$ is neglected) and electromagnetism from the antisymmetric part (which manifests itself as an $0(l^2)$ effect). It is also to be stressed that in this formulation, we are working with noncommutative effects at the Compton scale, this being true for the Weyl like formulation also.\\
\section{Introduction}
In an earlier communication \cite{ann}, based on a discrete spacetime noncommutative geometrical approach, we had shown that it was possible to reconcile electromagnetism and gravitation. It is of course well known that nearly ninety years of effort has gone in to get a unified description of electromagnetism and gravitation starting with Hermann Weyl's original Gauge Theory. It is only in the recent years that approaches in Quantum Gravity and Quantum Super Strings, amongst a few other theories are pointing the way to a reconciliation of these two forces. These latest theories discard the differentiable spacetime of earlier approaches and rely on a lattice like approach to spacetime, wherein there is a minimum fundamental interval which replaces the point space time of earlier theories. Indeed as Hooft has remarked, ``It is some what puzzling to the present author why the lattice structure of space and time had escaped attention from other investigators up till now....'' \cite{cu,hooft,lee} Infact we had recently shown that within this approach, it is possible to get a rationale for the de Broglie wavelength and Bohr-Sommerfeld quantization relations as well \cite{bgsafdb}. Nevertheless, the link with the gauge theories of other interactions, based as they are, on spin 1 particles, is not clear, because the graviton is a spin 2 particle (or alternatively, the gravitational metric is a tensor).
\section{A Gauge like Formulation}
In this latter context, we will now argue that it is possible for both electromagnetism and gravitation to emerge from a gauge like formulation. In Gauge Theory, which is a Quantum Mechanical generalization of Weyl's original geometry, we generalize, as is well known, the original phase transformations, which are global with the phase $\lambda$ being a constant, to local phase transformations with $\lambda$ being a function of the coordinates \cite{jacob}. As is well known this leads to a covariant gauge derivative. For example, the transformation arising from $(x^\mu) \to (x^\mu + dx^\mu)$,
\begin{equation}
\psi \to \psi e^{-\imath e \lambda}\label{e1}
\end{equation}
leads to the familiar electromagnetic potential
\begin{equation}
A_\mu \to A_\mu - \partial_\mu \lambda\label{e2}
\end{equation}
The above transformation, ofcourse, is a symmetry transformation. In the transition from (\ref{e1}) to (\ref{e2}), we expand the exponential, retaining terms only to the first order in coordinate differentials.\\
Let us now consider the case where there is a minimum cut off in the space time intervals. As is well known this leads to a noncommutative geometry (Cf.ref.\cite{ann})
\begin{equation}
[dx_\mu , dx_\nu ] = O(l^2)\label{e3}
\end{equation}
where $l$ is the minimum scale. From (\ref{e3}) it can be seen that if $O(l^2)$ is neglected, we are back with the familiar commutative spacetime. The new effects of fuzzy spacetime arise when the right side of (\ref{e3}) is not neglected. Based on this the author had argued that it is possible to reconcile electromagnetism and gravitation \cite{nc1,nc2,fpl1,fpl2}. If in the transition from (\ref{e1}) to (\ref{e2}) we retain, in view of (\ref{e3}), squares of differentials, in the expansion of the function $\lambda$ we will get terms like
\begin{equation}
\left\{ \partial_\mu \lambda \right\} dx^\mu + \left(\partial_\mu \partial_\nu + \partial_\nu \partial_\mu \right) \lambda \cdot dx^\mu dx^\nu\label{e4}
\end{equation}
where we should remember that in view of (\ref{e3}), the derivatives (or the product of coordinate differentails) do not commute. As in the usual theory the coefficient of $dx^\mu$ in the first term of (\ref{e4}) represents now, not the gauge term but the electromagnetic potential itself: Infact, in this noncommutative geometry, it can be shown that this electromagnetic potential reduces to the potential in Weyl's original gauge theory \cite{jacob,nc1}.\\
Without the noncommutativity, the potential $\partial_\mu \lambda$ would lead to a vanishing electromagnetic field. However Dirac pointed out in his famous monopole paper in 1930 that a non integrable phase $\lambda (x,y,z)$ leads as above directly to the electromagnetic potential, and moreover this was an alternative formulation of the original Weyl theory \cite{dirac2,nc3}.\\
Returning to (\ref{e4}) we identify the next coefficient with the metric tensor giving the gravitational field:
\begin{equation}
ds^2 = g_{\mu \nu} dx^\mu dx^\nu = \left(\partial_\mu \partial_\nu + \partial_\nu \partial_\mu \right) \lambda dx^\mu dx^\nu\label{e5}
\end{equation}
Infact one can easily verify that $ds^2$ of (\ref{e5}) is an invariant. We now specialize to the case of the linear theory in which squares and higher powers of  $h^{\alpha \beta}$ can be neglected. In this case it can easily be shown that
\begin{equation}
2 \Gamma^\beta_{\mu \nu} = h_{\beta  \mu ,\nu} + h_{\nu \beta ,\mu} - h_{\mu \nu ,\beta}\label{e6}
\end{equation}
where in (\ref{e6}), the $\Gamma$s denote Christofell symbols. From (\ref{e6}) by a contraction we have
\begin{equation}
2\Gamma^\mu_{\mu \nu} = h_{\mu \nu ,\mu} = h_{\mu \mu , \nu}\label{e7}
\end{equation}
If we use the well known gauge condition \cite{ohan}
$$\partial_\mu \left(h^{\mu \nu} - \frac{1}{2} \eta^{\mu \nu} h_{\mu \nu}\right) = 0, \, \mbox{where}\, h = h^\mu_\mu$$
then we get
\begin{equation}
\partial_\mu h_{\mu \nu} = \partial_\nu h^\mu_\mu = \partial_\nu h\label{e8}
\end{equation}
(\ref{e8}) shows that we can take the $\lambda$ in (\ref{e4}) as $\lambda = h$, both for the electromagnetic potential $A_\mu$ and the metric tensor $h_{\mu \nu}$. (\ref{e7}) further shows that the $A_\mu$ so defined becomes identical to Weyl's gauge invariant potential \cite{berg}.\\
However it is worth reiterating that in the present formulation, we have a noncommutative geometry, that is the derivatives do not commute and moreover we are working to the order where $l^2$ cannot be neglected. Given this condition both the electromagnetic potential and the gravitational potential are seen to follow from the gauge like theory. By retaining coordinate differential squares, we are even able to accommodate apart from the usual spin 1 gauge particles, also the spin 2 graviton which otherwise cannot be accommodated in the usual gauge theory. If however $O(l^2) = 0$, then we are back with commutative spacetime, that is a usual point spacetime and the usual gauge theory describing spin 1 particles.\\
We had reached this conclusion in ref.\cite{ann}, though from a different, nongauge point of view. The advantage of the present formulation is that it provides a transparent link with conventional theory on the one hand, and shows how the other interactions described by non Abelian gauge theories smoothly fit into the picture.\\
Finally it may be pointed out that the author had argued that a fuzzy spacetime input explains why the purely classical Kerr-Newman metric gives the purely Quantum Mechanical anomalous gyromagnetic ratio of the electron \cite{gravcos,fpl3}, thus providing a link between General Relativity and electromagnetism. This provides further support to the above considerations.

\end{document}